\documentstyle[12pt]{article}
\newcommand{\aaa}{{(a)}}
\newcommand{\bb}{{(b)}}

\newcommand{\wb}{\bar}

\newcommand{\be}{\begin{equation}}
\newcommand{\ee}{\end{equation}}
\newcommand{\ben}{\begin{eqnarray}\displaystyle}
\newcommand{\een}{\end{eqnarray}}
\newcommand{\refb}[1]{(\ref{#1})}
\newcommand{\p}{\partial}

\begin{document}

{}~ \hfill\vbox{\hbox{hep-th/9712150}\hbox{MRI-PHY/P971233}
}\break

\vskip 3.5cm

\centerline{\large \bf Black holes and Elementary String States
in}
\centerline{\large \bf N=2 Supersymmetric String Theories} 

\vspace*{6.0ex}

\centerline{\large \rm Ashoke Sen
\footnote{E-mail: sen@mri.ernet.in}}

\vspace*{1.5ex}

\centerline{\large \it Mehta Research Institute of Mathematics}
 \centerline{\large \it and Mathematical Physics}

\centerline{\large \it  Chhatnag Road, Jhoosi,
Allahabad 211019, INDIA}

\vspace*{4.5ex}

\centerline {\bf Abstract}

We compare the logarithm of the degeneracy of BPS
saturated elementary string 
states and the
string modified Bekenstein-Hawking entropy of the corresponding
black holes in N=2 supersymmetric heterotic string compactification to
four dimensions. 
As in the case of N=4 supersymmetric theory, the two results
match up to an overall undetermined numerical factor. We also
show that this undetermined numerical constant is identical in
the N=2 and N=4 supersymmetric theories, therby showing that the
agreement between the Bekenstein-Hawking entropy and the
microscopic entropy for N=2 theories does not require any new
identity, other than the one already required for the N=4 theory.
A similar result holds for type II string compactification as
well.

\vfill \eject

\baselineskip=18pt

\section{Introduction}

The idea of relating black holes to elementary string states
is quite old\cite{OLD,SUSS,DUF,MOREBH}. One particularly fascinating
similarity between black holes and the elementary string states
is the large amount of degeneracy of states of each of them. For
black holes the existence of this large degeneracy is deduced
indirectly from the large Bekenstein-Hawking entropy carried by a
macroscopic black hole. On the other hand, for elementary string
states, this large degeneracy is a consequence of the Hagedorn
spectrum of states in string theory.  Despite this qualitative 
similarity, the
naive attempt to compare the entropy of a Schwarzschild black
hole to the logarithm
of the degeneracy of elementary string states
runs into difficulty due to the fact that the former is
proportional to the square of the mass of the black hole,
whereas the later is proportional to the mass of the elementary
string states. It was suggested in \cite{SUSS} 
that this discrepancy
is due to the large mass renormalization suffered by the state
due to quantum corrections. This idea was further developed in
\cite{HORPOL,HKRS,MATHUR,YOUM}. Other attempts at providing a
microscopic explanation of the entropy of a Schwarzschild black
hole have been made in \cite{KAL,SFET}.

In supersymmetric string theories, there are a class of states,
known as BPS states, which do not suffer any mass
renormalization, and hence can avoid the difficulty
mentioned above\cite{WITOL}. This feature was exploited in \cite{SENOLD} to
compare the entropy of extremal electrically
charged BPS black holes in 
heterotic string theory compactified on $T^6$ with the
logarithm of the degeneracy of elementary string states carrying
the same charge. In this case, however, the difficulty arises
from that fact that the extremal electrically
charged black hole 
has vanishing area of the event horizon,
and is singular at the horizon. This leads to an apparent
contradiction since the microscopic calculation based on the
degeneracy of the elementary string states gives a non-vanishing
answer for the entropy. However, one should keep in mind that the
black hole solution that has vanishing area of the event horizon
is obtained by solving the low energy equations of motion of
string theory. These equations are valid far away from the horizon, 
but are expected to be modified near the horizon where the curvature
is large. Since for these black holes
the string coupling constant goes to zero near
the horizon, the relevant corrections to be taken into account
are the world-sheet $\sigma$-model corrections {\it but not
string loop corrections.} Although our
present technology does not allow us to calculate this correction
explicitly, a general scaling argument was used in \cite{SENOLD}
to determine the modified area of the
event horizon up to an overall numerical factor, which
could not be determined from the scaling argument. The result
agreed exactly with the microscopic entropy calculated from the
degeneracy of elementary string states up to this overall
numerical factor. This calculation was later generalized to
toroidally compactified heterotic string theory in higher
dimensions\cite{PEET}. Further support to the identification of 
extremal black holes with elementary string states was 
provided in \cite{KM,CMP,MW,DDMW,EMP} by comparing the scattering 
involving these 
two sets of states. Refs.\cite{CMP,DGHW} explicitly identified the
states responsible for black hole degeneracy by constructing classical 
solutions describing oscillations of the underlying microscopic string.
A somewhat different approach to this problem has been discussed
in \cite{SUZ}.

Since then great progress has been made
towards understanding the microscopic
origin of black hole entropy by comparing the
Bekenstein-Hawking entropy to the microscopic entropy of
appropriate configuration of D-branes\cite{POLC} carrying the
same charge as the black hole\cite{STVA}. The main
advantage of using D-branes instead of elementary string states
is that for an appropriate configuration of D-branes, the event
horizon of the
corresponding black hole is non-singular and has finite area.
Thus the Bekenstein-Hawking entropy for these black holes can be
calculated unambiguously, and can be compared with the
corresponding microscopic answer obtained from the counting
of states of the D-brane.
The two calculations turn out to be
in exact agreement, including the
overall numerical factor. Since then this calculation has been
generalized to many other classes of
black holes\cite{MALREV}, including black 
holes in N=2 supersymmetric string 
theory\cite{FKS,STRO,KLMS,MSW,VAFA,FK,SH,BLKLM,BM,MAL}.
Much progress has also been made in extending this analysis to
slightly non-extremal black holes\cite{NONEX} and in providing a 
microscopic explanation of 
Hawking radiation from these black holes\cite{DMW,DASMAT,MALSTR}. An
alternative derivation 
of the microscopic entropy of these black
holes has also been attempted\cite{LARWIL,CVTS,TSTWO}
by mapping this problem to that of counting of elementary
string states.

Although these developments mark tremendous progress in our
understanding of microscopic description of black holes, they do
not directly address the problem of
studying the relationship between elementary string
states and black holes.
The results of \cite{SENOLD,PEET} provide strong
evidence for such a relationship, since here an explanation for
the Bekenstein-Hawking entropy of the black hole is provided in
terms of the degeneracy of the microscopic states of the
elementary string excitations. In particular, if one could take
into account the higher derivative terms in the string effective
action to compute the overall numerical constant in the
expression for the Bekenstein-Hawking entropy and show that the
result agrees with the microscopic entropy computed from the
degeneracy of string states, it would put the correspondence on
a much firmer ground. However, as we have already mentioned, our
current technology in string theory does not permit us to do this
computation. Another direction could be in extending these 
results to non-supersymmetric black holes. Some progress
in this direction has been made in \cite{JM,CY,DLR,DAB,DMR}. 

In this paper
we shall generalise the result of \cite{SENOLD} to more general
string theories in four dimensions. Since for the
comparison to be meaningful we need BPS states, we need the four
dimensional theory to have at least N=2 supersymmetry.  
We shall focus our attention on the most general N=2 string
compactification\footnote{Although for N=2 supersymmetric
theories the degeneracy of string states can jump discontinuously
as we move in the moduli space\cite{SEIWIT}, we shall work at
weak string coupling where such phenomena are not expected to
occur.}, but as we shall see, based on this analysis
we shall also be able to make some general remark about more
general N=4 string theories other than the ones obtained by
toroidal compactification of heterotic string theory\cite{NARAIN}.

There are several ways to get N=2 supersymmetric string
theories in four
dimensions. One class of these theories is obtained by heterotic
string compactification on a conformal field theory with (0,4)
world-sheet supersymmetry. The simplest example is heterotic
string theory on $K3\times T^2$, but more general
compactifications based on asymmetric orbifolds, and other exotic
conformal field theories are possible\cite{BLFM,BANDIX}. This is the
class of theories on which we shall focus most of our attention.
We can also get N=2 supersymmetric string theories from
compactifying type II string theory on Calabi-Yau manifolds. But
in these theories all gauge fields arise from the Ramond-Ramond
sector, and as a result, none of the elementary string states are
BPS states since they do not carry any Ramond-Ramond charge.
Finally, we can also get N=2 supersymmetric string theories from
asymmetric compactification of type II theories where all the
space-time supersymmetries come from the right (or left) moving
sector on the world-sheet\cite{ASYM}. These theories do have BPS 
saturated elementary
string states. As we shall see, the analysis for these theories is
very similar to that for the heterotic string theories.\footnote{One 
could also consider compactified type I theory, but the spectrum of 
elementary string states in this theory only has a limited number of
BPS states associated with states carrying internal momentum. In
particular, there are no BPS elementary string states
with large degeneracy.}

We shall show that for the most general N=2 heterotic string
compactification, the Bekenstein-Hawking entropy of an extremal
electrically charged black hole agrees with the microscopic
entropy calculated from the degeneracy of elementary string
states up to an overall undetermined numerical factor, in a manner
identical to that in the case of toroidally compactified
heterotic string theory. More importantly, we shall show that the
computation of this overall numerical factor in the N=2 theories
is independent of the details of the
compactification and is identical 
to that in the case of toroidally compactified heterotic
string theory. In other words if one is able to compute this
factor for the toroidally compactified heterotic string theory
and show precise agreement between the Bekenstein-Hawking entropy
and the microscopic entropy for this theory, then it also
guarantees precise agreement bewtween the two entropies in all
N=2 supersymmetric heterotic compactification. The same result
holds for non-toroidal N=4 supersymmetric heterotic string
compactification discussed in \cite{CHL}. 

A similar story is repeated for type II string compactification. For
any four dimensional type II string compactification with at least 
two supersymmetries coming froim the right-moving sector of the
world-sheet, $-$
toroidal
compactification with N=8 supersymmetry, symmetric or asymmetric
compactification with N=4 supersymmetry, or asymmetric N=2
supersymmetric compactification $-$ the microscopic entropy
associated with elementary BPS saturated type II string states
agrees with the Bekenstein-Hawking entropy of the corresponding
extremal black hole up to an overall numerical constant. This
constant is again identical in all of these type II string
compactifications. However {\it it is different from the
corresponding constant in the heterotic string compactification}.
This is just as well, since the required value of this constant
for exact agreement between the two entropies is different in
type II and heterotic string theories, due to different pattern
of growth of the number of states with energy in the two theories.

\section{Black Hole Entropy}

At tree level, the massless bosonic field content in the
most general heterotic string compactification to four
dimensions with N=2 supersymmetry 
at a generic point in the
moduli space is the metric $g_{\mu\nu}$,
$(r+2)$ U(1) gauge fields $A_\mu^\aaa$
($0\le\mu\le 3$, $1\le a\le (r+2)$) and $2r+2$ scalars
locally parametrizing the moduli
space\cite{FKP,CLO,CLM,WKLL,AFGNT,FRE}:
\be \label{e1}
{SO(2,r)\over SO(2)\times SO(r)}\times {SL(2,R)\over U(1)}\, .
\ee
The scalars labelling the first coset may be represented by an
$(r+2)\times (r+2)$ matrix $M$ satisfying,
\be \label{e2}
M^T=M, \qquad MLM^T = L\, ,
\ee
where
\be \label{e3}
L = \pmatrix{I_2 & \cr & -I_r}\, .
\ee
$I_n$ denotes $n\times n$ identity matrix. The second coset is
labelled by a complex scalar $\lambda$ taking value in the upper
half plane:
\be \label{e4}
\lambda \equiv \lambda_1 + i\lambda_2 = a + ie^{-\Phi}\, ,
\ee
where $a$ is the axion field obtained by dualizing the rank two
antisymmetric tensor field, and $\Phi$ is the dilaton field.
This moduli space gets modified by quantum
corrections\cite{WKLL,AFGNT}, but since the string coupling
vanishes at the black hole horizon, we shall not need to consider
these corrections.
There may also be massless scalars belonging to the
hypermultiplet but they will play no role in our discussion, 
since in constructing the black hole solution we shall set these
hypermultiplet fields to zero. The
tree level low energy effective action
involving the metric and the vector multiplet fields $A_\mu^\aaa$,
$M$ and $\lambda$ is given by,
\ben \label{e5}
S &=& {1\over 32\pi} \int d^4 x \sqrt{-g} \Big[ R - g^{\mu\nu}
{\p_\mu\lambda\p_\nu \bar \lambda\over 2(\lambda_2)^2} + {1\over
8} g^{\mu\nu} Tr(\p_\mu M L \p_\nu M L) \nonumber \\
&& -\lambda_2 g^{\mu\mu'} g^{\nu\nu'} F^\aaa_{\mu\nu} (LML)_{ab}
F^\bb_{\mu'\nu'} + {1\over 2} \lambda_1 (\sqrt-g)^{-1} \
\epsilon^{\mu\nu\rho\sigma} F^\aaa_{\mu\nu} L_{ab}
F^\bb_{\rho\sigma}\Big]\, ,
\een
where $F^\aaa_{\mu\nu}$ is the field strength associated with 
$A_\mu^\aaa$ and $R$ is the Ricci scalar. Our normalization 
conventions are identical to those in \cite{SENOLD,SREV}.
This action is invariant under the $SO(2,r)$ transformation:
\be \label{e7}
M\to \Omega M \Omega^T, \quad A^\aaa_{\mu}\to
\Omega_{ab}A^\bb_{\mu}, \quad \lambda\to \lambda, \quad
g_{\mu\nu}\to g_{\mu\nu}\, ,
\ee
where $\Omega$ is an $(r+2)\times (r+2)$ matrix satisfying 
\be \label{e7a}
\Omega L \Omega^T = L\, .
\ee
The equations of motion are also invariant under an SL(2,R)
transformation\cite{SREV}. 

The form of the action \refb{e5} is
dictated to a large extent by the local $N=2$ supersymmetry of the 
theory\cite{FRE}. 
Note that this action is the restriction of the
$SO(6,22)$ invariant effective action\cite{MAHSCH,SREV} for
heterotic string theory on $T^6$ to a subspace where the
$SO(6,22)$ matrix $M$ is restricted to take value in an $SO(2,r)$
subgroup, and
the 28 gauge fields in the vector representation of $SO(6,22)$ are
restricted to lie in the subspace transforming in the vector
representation of $SO(2,r)$. 
However, there is a much more powerful result relating these two
effective actions that we shall need. In general both, the action
of the toroidally compactified heterotic theory and that of the N=2
supersymmetric theory, will receive higher derivative corrections even at
the string tree level.
These can be viewed as coming from higher loop
contribution to the $\beta$-function of the corresponding
$\sigma$-model\cite{SBETA,CFMP}, or, equivalently,
from order $\alpha'$ and higher order
corrections to the tree level string $S$-matrix elements.
It turns out that even after these 
higher derivative terms are included, the {\it tree level} effective
action involving the metric and the vector multiplet
fields of the N=2
supersymmetric heterotic string theory is given by the
restriction of the tree level effective action of the corrsponding
toroidally compactified heterotic string theory.

This can be seen as follows. The $(r+2)$ vector fields in the N=2
theory are associated with two right-moving $U(1)$ super-currents
$(j^\alpha(z), J^\alpha(z))$ ($1\le\alpha\le 2$)
and $r$ left moving $U(1)$
currents $\wb J^i(\wb z)$ ($3\le i\le
(r+2)$).\footnote{In our convention the world-sheet and hence
space-time supersymmetry is in the right-hand sector. Also the
right-handed currents are holomorphic.} Here $j^\alpha$, 
$J^\alpha$ and $\wb J^i$
have dimensions $(0,{1\over 2})$, $(0,1)$ and $(1,0)$
respectively. Let $X^\mu$ and $\psi^\mu$ denote
respectively the bosonic
coordinates and their right-handed superpartners associated with
the non-compact directions. Then the vertex operators of the
$(r+2)$ vector fields in the $-1$ representation\cite{FMS,KNIZ}
are given by\cite{BANDIX,LLT,WKLL}:
\ben \label{e10}
V^{\alpha\mu}_{-1} &=& e^{-\phi} j^\alpha(z) \wb\p X^\mu(\wb z)
e^{ik\cdot X}\, , \nonumber \\
V^{i\mu}_{-1} &=& e^{-\phi} \psi^\mu(z) \wb J^i(\wb z) e^{ik\cdot
X}\, , 
\een
where $\phi$ is the bosonized super-ghost. The same vertex operators in
the 0 representation are given by
\ben \label{e11}
V^{\alpha\mu}_{0} &=& J^\alpha(z) \wb\p X^\mu(\wb z)
e^{ik\cdot X}\, , \nonumber \\
V^{i\mu}_{0} &=& \Big(\p X^\mu(z) + i k\cdot \psi(z)\psi^\mu(z)\Big) 
\wb J^i(\wb z) e^{ik\cdot
X}\,  
\een
The vertex operators of the $2r$ scalars parametrizing the coset
$SO(2,r)/(SO(2)\times SO(r))$ in the $-1$ picture are:
\be \label{e12}
V^{\alpha i}_{-1} = e^{-\phi} j^\alpha(z) \wb J^i(\wb z)
e^{ik\cdot X}\, .
\ee
The same vertex operators in the 0 picture are:
\be \label{e13}
V^{\alpha i}_{0} = \Big(J^\alpha(z) + i k\cdot \psi(z)
j^\alpha(z)\Big) \wb J^i(\wb z)
e^{ik\cdot X}\, .
\ee
Finally the vertex operators for the metric $g_{\mu\nu}$ and the
scalars $\lambda$ in both $-1$ and $0$ picture are constructed
solely in terms of the world-sheet fields $X^\mu$, $\psi^\mu$ and
their derivatives, and the (super-)ghost fields.

The tree level S-matrix involving the metric and the particles in
the vector multiplet are computed from the correlation functions
of various combinations of these vertex operators on the sphere.
As can be seen from the structure of the vertex operators, these
correlation functions are completely independent of the details
of the conformal field theory describing this compactification.
The correlators of operators constructed from $X$ and $\psi$ of
course are manifestly independent of this conformal field theory.
But on the sphere, even the correlators of $j^\alpha$, $J^\alpha$
and $\wb J^i$ are determined solely from the $U(1)$ (super-)
current algebra and hence are insensitive to the details of the
conformal field theory. Note that this argument does not assume that
these (super-) currents are decoupled from the rest of the
conformal field theory. In particular there may be non-trivial
correlation between the U(1) charges carried by a state and 
the vertex operator representing it
in the rest of the conformal field theory. These
states affect the correlation function of the currents on the
torus and the higher genus Riemann surfaces, but not on the
sphere.

Note in particular that vertex operators with identical
structure and correlation functions exist in toroidally
compactified heterotic string theory as well.
In heterotic string theory on $T^6$, we can
take the right-moving $U(1)$ supercurrents $(j^\alpha, J^\alpha)$ to be
two of the six right-moving supercurrents, 
and the left-moving $U(1)$ currents $\wb
J^i$ to be the $r$ of the 22 left-moving 
$U(1)$ currents in this world-sheet theory. The
$S$-matrix elements and hence the effective action involving
these restricted set of
fields will be insensitive to the fact that we now have a
toroidal compactification instead of $N=2$ supersymmetric
compactification, and hence will yield the same answer. 
This shows that in an $N=2$ supersymmetric heterotic compactification,
the tree level $S$-matrix, and hence the
effective action involving
the metric and the vector multiplet
fields is independent of the choice
of the internal conformal field theory, and is given by an 
appropriate restriction of the full tree level effective action
of the toroidally compactified heterotic string theory.

Given this result, we proceed as follows.
Since the effective action of the N=2 theory is the restriction
of the effective action of the N=4 theory, the electrically
charged black hole solutions in the N=2 theory are simply the
corresponding black hole solutions in the N=4 theory with
appropriate restriction on the charge. In this case since there
are $(2+r)$ $U(1)$ gauge fields, we have a $(2+r)$ dimensional
charge vector $Q^\aaa$. As in \cite{SENOLD} we define the left and
the right-handed components of the charge as:
\ben \label{e14}
Q_R^2 &=& {1\over 2} Q^\aaa (L\langle M\rangle L + L)_{ab} Q^\bb\,
, \nonumber \\
Q_L^2 &=& {1\over 2} Q^\aaa (L\langle M\rangle L - L)_{ab} 
Q^\bb\, , 
\een
where $\langle~\rangle$ denotes 
the asymptotic value of a field in the presence of a black hole.
The BPS condition requires\footnote{This condition is modified by
string loop corrections, but we shall work at weak coupling where
this relation is valid.}:
\be \label{e15}
m^2 = Q_R^2/8g^2\, ,
\ee
where $m$ is the ADM mass of the black hole, and $g$ is the
string coupling constant. Arguments identical to that in
ref.\cite{SENOLD} give the following expression for the string
modified Bekenstein-Hawking entropy of the black hole:
\be \label{e16}
S_{BH} = {2\pi C\over g} \sqrt{m^2 - {Q_L^2\over 8g^2}}\, .
\ee
Here $C$ is the undermined numerical constant alluded to before.
The computation of $C$ involves the higher derivative terms in
the effective action {\it at the string tree level}, but not
string loop corrections, since the string coupling constant
vanishes as we approach the horizon. Since these higher
derivative terms are identical to the ones in the toroidally
compactified heterotic string theory, we conclude that the
numerical constant $C$ must also be identical to that in
heterotic string theory on $T^6$.

We now need to compute the microscopic entropy by counting the
degeneracy of the BPS saturated elementary string states with the
same quantum number. In the normalization convention used in
\cite{SREV,SENOLD} the NS sector mass formula for an elementary string 
states carrying charge $Q$ is given by:
\be \label{e17}
m^2 = {Q_R^2\over 8g^2} = {g^2\over 8} \Big( {Q_L^2\over g^4}
+ 2\Delta -2\Big)\, ,
\ee
where $\Delta$ represents the matter contribution to $\wb L_0$ 
other than the contribution due to the zero modes of $\wb J^i$,
$\wb\p X^\mu$.
(In a toroidal compactification $\Delta$ would represent
the oscillator contribution to $\wb L_0$). We need to calculate
the degeneracy $d$ of such states for large $\Delta$. This can be
done following the procedure developed in 
\cite{CARDY,KAVA,HARMOR,STVA} and the
answer is:
\be \label{e18}
d \simeq \exp\Big(2\pi \sqrt{c_L\Delta\over 6}\Big)\, ,
\ee
where $c_L$ is the total central charge of the left-handed part
of the conformal field theory. If we work in the light-cone gauge
so that all states are physical, we have
\be \label{e19}
c_L =24\, .
\ee
This gives the following expression for the microscopic entropy:
\be \label{e20}
S_{micro} \equiv \ln d \simeq 4\pi \sqrt\Delta = {8\pi\over g}
\sqrt{m^2 -{Q_L^2\over 8g^2}}\, ,
\ee
where we have used eq.\refb{e17} and the approximation $\Delta >>
1$.
This is again identical to the expression for the microscopic
entropy in heterotic string theory on $T^6$. \refb{e20} and
\refb{e16} agree if,
\be \label{e21}
C= 4\, .
\ee
This is the same condition found in \cite{SENOLD} for the
agreement between the Bekenstein-Hawking entropy and the 
microscopic entropy for  heterotic string theory on $T^6$. Thus
we see that the criteria that guarantees the agreement between
microscopic and the Bekenstein-Hawking entropy for electrically
charged black holes in heterotic string theory on $T^6$ also
guarantees similar agreement for N=2 supersymmetric
compactification of the heterotic string theory.

{}From this discussion it is clear that the same analysis can 
also be extended to more general N=4 supersymmetric
compactifications of the heterotic string theory described in
\cite{CHL}. The relevant part of the
tree level effective action in these theories
will be given by an appropriate restriction of the  tree level
effective action of heterotic string theory on $T^6$. Thus the
Bekenstein-Hawking entropy of these black holes will again be 
given by \refb{e16} with the same constant $C$. Furthermore the
microscopic entropy computed from the
degeneracy of BPS states with a given set of charges is also
given by the same formula \refb{e20}. Thus demanding the
agreement between the microscopic entropy and the
Bekenstein-Hawking entropy gives us back the same equation
\refb{e21}.

It is now also clear how to extend this analysis to type II
compactification. In this case the `parent' theory will be type
II string theory on $T^6$. Since the elementary string states carry
only NS sector charges, we can restrict our attention to the
effective action involving NS sector bosonic fields only. The low
energy effective action is given by \refb{e5} with $M$
representing
an $O(6,6)$ matrix. There are higher derivative corrections to
this action at the string tree level. 
These corrections are different from those
in the heterotic string theory, since in this case there is local
world-sheet supersymmetry in both the left and the right sector,
and hence the vertex operators are constructed from superfields
in both these sectors. This would modify the answer for the
Bekenstein-Hawking entropy for an electrically chaged black hole
to\footnote{Here we are considering states which break all 
space-time supersymmetries
coming from the left-moving sector of the world-sheet, and half of the
supersymmetries coming from the right-moving sector.}
\be \label{e22}
S_{BH} = {2\pi C'\over g}\sqrt{m^2 -{Q_L^2\over 8g^2}}\, ,
\ee
where $C'$ is a new numerical constant. This is just as well,
since the computation of the microscopic entropy in this
theory also differs from that in the heterotic string theory.
In particular,
$c_L$ appearing in eq.\refb{e18} now takes the value
\be \label{e19a}
c_L = 12\, ,
\ee
reflecting the contribution from the eight bosonic and eight
fermionic world-sheet fields in the light-cone gauge. This gives,
\be \label{e23}
S_{micro} = {4\sqrt 2 \pi\over g} \sqrt{m^2 - {Q_L^2\over
8g^2}}\, .
\ee
Agreement between \refb{e22} and \refb{e23} now requires
\be \label{e24}
C' = 2\sqrt 2 \, .
\ee
Consider now any other compactification of the type II theory.
Following the same argument as in the heterotic case, we see
that the full tree level
effective action involving the metric, and the NS sector
vectors and scalars constructed from the $U(1)$ super-currents 
will be given by a restriction of the tree level
effective action of type
IIB on $T^6$. Thus the Bekenstein-Hawking entropy of an
electrically charged BPS black hole will be given by an
expression identical to \refb{e22}, with the same constant $C'$.
On the other hand the computation of the microscopic entropy also
proceeds in an identical manner, with $c_L$ given by 12 in each
case. Thus the same equation \refb{e24} guarantees the agreement
between the Bekenstein-Hawking entropy and the microscopic
entropy in all cases.

\section{Conclusion}

In this paper we have generalised the analysis of 
refs.\cite{SENOLD,PEET} to black holes in more general heterotic and
type II string theories in four dimensions with at least N=2
supersymmetry. We have shown that in each case, the black hole
entropy calculated from the string modified area of the event horizon
agrees with the microscopic entropy calculated from the degeneracy of
elementary string states, up to an overall numerical factor. 
Furthermore, we find that this numerical constant is universal in all
heterotic string compactifications, and also in all type II 
compactifications. This means that if this constant has the correct
value in one heterotic compactification, then it 
will automatically have the correct value for all heterotic string
compactification. A similar result holds for type II compactification.

The explicit computation of this constant remains a problem for the future.
The relevant $\sigma$-models describing the conformal field theory near
the event horizon are the chiral null models discussed in \cite{HORTSE}.
But in order to compute the coefficient from these chiral null models
one needs an abstract definition of the area of the event horizon in terms 
of the two dimensional conformal field theory describing string propagation
in a black hole background. Such a definition is lacking at this moment.

In particular, there is a specific choice of metric such that the area of
the event horizon, measured in this metric, remains zero to all orders in
the $\sigma$-model loop expansion\cite{HORTSE}. 
But there is no reason to believe
that this is the metric that we should be using in computing the area of
the event horizon while calculating the entropy. On the other hand, if we
insist on working with this metric, then there is no reason to believe
that the relationship between the entropy and the area of the event
horizon does not get modified by higher loop terms in the $\sigma$-model.

\end{document}